\documentclass[sort&compress,
    final            
  4apaper]
  {aipproc}

\layoutstyle{8x11single}

    \usepackage{bm}
\usepackage{color}

\newcommand\beq{\begin{equation}}
\newcommand\eeq{\end{equation}}
\newcommand\beqa{\begin{eqnarray}}
\newcommand\eeqa{\end{eqnarray}}
\newcommand{\nn}{\nonumber\\}

\newcommand{\text}{\mathrm}

\newcommand{\rr}{\mathbf{r}}

\newcommand{\cc}{\mathbf{v}}

\newcommand{\ww}{\bm{\omega}}

\newcommand{\q}{\kappa}

\newcommand{\Tt}{T^\text{tr}}

\newcommand{\Tr}{T^\text{rot}}
\newcommand{\Trb}{\overline{T}^\text{rot}}

\newcommand{\zt}{\xi^\text{tr}}

\newcommand{\zr}{\xi^\text{rot}}

\newcommand{\tr}{\text{tr}}
\newcommand{\rot}{\text{rot}}

\newcommand{\kk}{\bm{\mathsf{k}}}

\begin{document}

\title{Relative Entropy  of a Freely Cooling Granular Gas}

 \classification{45.70.Mg, 
 05.20.Dd, 
  51.10.+y, 
  83.80.Fg} 
 \keywords{Boltzmann equation,  Granular gases, Rough spheres, Entropy, Cumulants
 }

\author{Andr\'es Santos}{
  address={Departamento de F\'{\i}sica, Universidad de
Extremadura, E-06071 Badajoz, Spain} }

\author{Gilberto M. Kremer}{
  address={Departamento de F\'{\i}sica, Universidade Federal do Paran\'a, Curitiba, Brazil} }

\begin{abstract}
The time evolution and stationary values of the  entropy per particle of a homogeneous freely cooling granular gas, relative to the maximum entropy consistent with the instantaneous translational and rotational temperatures, is analyzed by means of a Sonine approximation involving fourth-degree cumulants. The results show a rich variety of dependencies of the relative entropy on time and on the coefficients of normal and tangential restitution,  including a peculiar behavior in the quasi-smooth limit.
\end{abstract}

\maketitle


\section{Introduction and basic equations}
The dynamics of a dilute granular gas, modeled  as a system of hard spheres colliding inelastically with constant coefficients of normal ($\alpha$) and tangential ($\beta$) restitution, can be described at a mesoscopic level by the (inelastic) Boltzmann equation for the one-body velocity distribution function $f(\mathbf{r},\mathbf{v},\ww,t)$ \cite{JR85,LS87,GS95,HZ97,LHMZ98,ZVPSH98,GNB05,Z06,BPKZ07,KBPZ09,SKG10,S11a,S11b,SKS11}:
\beq
\partial_t f(\rr,\cc,\ww,t)+\cc\cdot \nabla f(\rr,\cc,\ww,t)=J[\cc,\ww|f(\rr,\cdot,\cdot,t),f(\rr,\cdot,\cdot,t)],
\label{2.1}
\eeq
where the collision operator is
\beq
J\left[{\bf v}_{1},\ww_1|f,f\right]=\sigma^{2} \int
d{\bf v}_{2}\int d\ww_2\int d\widehat{\bm{\sigma}} \, \Theta\left(\mathbf{g}\cdot\widehat{\bm{\sigma}}\right)\left(\mathbf{g}\cdot\widehat{\bm{\sigma}}\right)
\left[
(\alpha\beta)^{-2}f({\bf v}_{1}'',\ww_1'')f({\bf v}_{2}'',\ww_2'')-f({\bf v}_{1},\ww_1)f({\bf
v}_{2},\ww_2)\right].
\label{2.2}
\eeq
Here, $\sigma$ is the diameter of a sphere, $\Theta(x)$ is Heaviside's step function, $\widehat{\boldsymbol{\sigma}}$ is a unit vector
directed along the centers of the two colliding particles, $\mathbf{g}=\mathbf{v}_1-\mathbf{v_2}$ is the relative velocity of the center of masses, and the double primes denote pre-collisional velocities, i.e., $(\cc_{1,2}'',\ww_{1,2}'')\stackrel{\text{coll}}{\longrightarrow}(\cc_{1,2},\ww_{1,2})$.
Conservation of linear and angular momenta implies
\beq
\cc_1''+\cc_2''=\cc_1+\cc_2,
\label{cons1}
\eeq
\beq
 I\ww_1''-m\frac{\sigma}{2}\widehat{\bm{\sigma}}\times\cc_1''=I\ww_1-m\frac{\sigma}{2}\widehat{\bm{\sigma}}\times\cc_1
,\quad I\ww_2''+m\frac{\sigma}{2}\widehat{\bm{\sigma}}\times\cc_2''=I\ww_2+m\frac{\sigma}{2}\widehat{\bm{\sigma}}\times\cc_2,
\label{cons2}
\eeq
where $m$ and $I$ are the mass and the moment of inertia of a sphere, respectively. The pre- and post-collisional relative velocities of the two points \emph{at contact} are $\overline{\mathbf{g}}''=\mathbf{g}''-\frac{\sigma}{2}\widehat{\bm{\sigma}}\times(\ww_1''+\ww_2'')$ and $\overline{\mathbf{g}}=\mathbf{g}-\frac{\sigma}{2}\widehat{\bm{\sigma}}\times(\ww_1+\ww_2)$, respectively. The coefficients of restitution $\alpha$ and $\beta$ relate the normal and tangential components of $\overline{\mathbf{g}}$ to those of $\overline{\mathbf{g}}''$:
\beq
\widehat{\bm{\sigma}}\cdot \overline{\mathbf{g}}=-\alpha \widehat{\bm{\sigma}}\cdot \overline{\mathbf{g}}'',\quad
\widehat{\bm{\sigma}}\times \overline{\mathbf{g}}=-\beta \widehat{\bm{\sigma}}\times\overline{\mathbf{g}}''.
\label{restcoeff}
\eeq
The coefficient $\alpha$ ranges from $\alpha=0$ (collisions perfectly inelastic) to $\alpha=1$ (collisions perfectly elastic), while the coefficient $\beta$ ranges from $\beta=-1$ (spheres perfectly smooth) to $\beta=1$ (spheres perfectly rough). Except if $\alpha=1$ and $\beta=\pm 1$, energy is dissipated upon collisions. Using Eqs.\ \eqref{cons1}--\eqref{restcoeff}, it is possible to obtain the (restituting) collision rules  \cite{Z06,SKG10,S11a,S11b,SKS11}
\beq
\label{2.3a}
{\bf v}_{1}''={\bf v}_{1}-\mathbf{C}, \quad
{\bf v}_{2}''={\bf
v}_{2}+\mathbf{C}
\eeq
\beq
\label{2.3b}
\ww_1''=\ww_1-\frac{2}{\sigma \kappa}\widehat{\bm{\sigma}}\times \mathbf{C},\quad \ww_2''=\ww_2-\frac{2}{\sigma \kappa}\widehat{\bm{\sigma}}\times \mathbf{C},
\eeq
where $\kappa\equiv {4I}/{m\sigma^2}$ is the reduced moment of inertia and we have called
\beq
\mathbf{C}\equiv
\frac{1+\alpha
^{-1}}{2}(\widehat{\boldsymbol{\sigma}}\cdot {\bf
g})\widehat{\boldsymbol {\sigma}}+\frac{\kappa}{1+\kappa}\frac{1+\beta^{-1}}{2}\left[\mathbf{g}-(\widehat{\boldsymbol{\sigma}}\cdot {\bf
g})\widehat{\boldsymbol {\sigma}}-\frac{\sigma}{2}\widehat{\bm{\sigma}}\times\left(\ww_1+\ww_2\right)\right].
\label{2.4}
\eeq

The basic velocity moments are the number density $n(\rr,t)$, the flow velocity $\mathbf{u}(\rr,t)$, the average spin (or mean angular velocity) $\bm{\Omega}(\rr,t)$, the granular translational temperature $T^\tr(\rr,t)$, and the granular rotational temperature $T^\rot(\rr,t)$. They are defined as
\beq
n(\rr,t)=\int d\cc\int d\ww\, f(\rr,\cc,\ww,t), 
\eeq
\beq
\mathbf{u}(\rr,t)=\langle \cc \rangle
, \quad \bm{\Omega}(\rr,t)=\langle \ww \rangle,
\eeq
\beq
T^\tr(\rr,t)=\frac{m}{3}\langle \left[\cc-\mathbf{u}(\rr,t)\right]^2 \rangle,\quad
T^\rot(\rr,t)=\frac{I}{3}\langle \omega^2 \rangle,
\eeq
where  we have introduced the short-hand notation
\beq
\langle \psi(\rr,\cc,\ww,t)\rangle\equiv \frac{1}{n(\rr,t)}\int d\cc\int d\ww\, \psi(\rr,\cc,\ww,t) f(\rr,\cc,\ww,t).
\eeq
As usually done in the literature on granular gases, the Boltzmann constant is absorbed in the definition of $\Tt$ and $\Tr$, so that these quantities have dimensions of energy.
Conservation of mass and linear momentum imply that 
\beq
\int d\cc\int d\ww\, J[\cc,\ww|f,f]=\int d\cc\int d\ww\, \cc J[\cc,\ww|f,f]=0.
 \eeq
On the other hand, neither the total angular velocity nor the translational or rotational kinetic energies are in general conserved by collisions:
\beq
\left.\frac{\partial \bm{\Omega}}{\partial t}\right|_{\text{coll}}\equiv\frac{1}{n} \int d\cc\int d\ww\, \ww J[\cc,\ww|f,f]=-\bm{\Lambda},
\label{2.9}
\eeq
\beq
\left.\frac{\partial \Tt}{\partial t}\right|_{\text{coll}}\equiv\frac{m}{3n}\int d\cc \int d\ww\,\cc^2 J[\cc,\ww|f,f]=-\zt\Tt,
\label{2.6t}
\eeq
\beq
\left.\frac{\partial \Tr}{\partial t}\right|_{\text{coll}}\equiv\frac{I}{3n}\int d\cc \int d\ww\,\omega^2 J[\cc,\ww|f,f]=-\zr \Tr.
\label{2.6r}
\eeq
The above equations define the spin production vector $\bm{\Lambda}$ and the energy production rates $\zt$ and $\zr$. While $\zt$ and $\zr$ can be positive or negative,  the net cooling rate $\zeta=(T^\tr\zt+T^\rot\zr)/(T^\tr+T^\rot)$ is positive definite \cite{SKS11}.

{In general, the collisional processes of particles with internal degrees of freedom are characterized by an energy  transfer between the translational and internal degrees of freedom. Here we are interested in the rotational degrees of freedom of  inelastic rough hard spheres and for that reason we have introduced separate translational and rotational temperatures, which are associated with their corresponding energies. The introduction of these two temperatures is important since, not only they exhibit different relaxation rates, but they characterize the non-equipartition of energy inherent to granular fluids \cite{LHMZ98,S11b}.}

A granular gas is intrinsically out of equilibrium and thus, in contrast to the case of energy conservation, the evolution of $f$ does not   obey an H theorem \cite{C88,GS03,K10a}, even if the system is isolated. On the other hand, it is worthwhile introducing the Boltzmann entropy density $n(\rr,t)s(\rr,t)$ \cite{K10b}, where
\beq
s(\rr,t)=- \left\langle\ln \frac{f(\rr,\mathbf{v},\ww,t)}{K}\right\rangle
\label{7}
\eeq
is the entropy per particle, $K$ being an irrelevant constant with the same dimensions as $f$.
Although $n(\rr,t)s(\rr,t)$ does not  qualify as a Lyapunov function in the inelastic case, its introduction is justified by information-theory arguments \cite{SW71} and also to make contact with the Boltzmann entropy density of a conventional gas \cite{K10a}.

{The connection between Shannon entropy in information theory and the one in statistical mechanics is discussed by Jaynes  \cite{J57}. The Shannon entropy of a discrete random variable which assumes the possible values $\{x_1,\dots, x_n\}$ with respective probabilities $\{p_1,\ldots,p_n\}$ is given by $H(\{p_i\})=-k\sum_ip_i\ln p_i$, where $k$ is a positive constant. A similar  expression for the entropy of a system in contact with a heat bath  holds, where $k$ is identified with Boltzmann's constant and $p_i$ with the probability to find the system in a certain microstate with energy $E_i$.}

Given local and instantaneous values of the quantities $n(\rr,t)$, $\mathbf{u}(\rr,t)$, $\bm{\Omega}(\rr,t)$, $T^\tr(\rr,t)$, and $T^\rot(\rr,t)$, the velocity distribution function that \emph{maximizes} the entropy density is the two-temperature Maxwellian
\beq
f_0(\cc,\ww)= n\left(\frac{m I}{4\pi^2 \Tt\Trb}\right)^{3/2}
\exp\left[-\frac{m\left(\cc-\mathbf{u}\right)^2}{2\Tt}-
\frac{I\left(\ww-\bm{\Omega}\right)^2}{2\Trb}\right],
\label{2.18}
\eeq
where 
\beq
\Trb\equiv\frac{I}{3n}\langle \left(\bm{\omega}-\bm{\Omega}\right)^2\rangle=\Tr-\frac{I\Omega^2}{3}
\eeq
is an alternative rotational temperature.
The  entropy  per particle associated with the distribution \eqref{2.18} is
\beq
s_0=3-\ln\left[\frac{n}{K}\left(\frac{m I}{4\pi^2 \Tt\Trb}\right)^{3/2}\right],
\label{s0}
\eeq
where we have taken into account that  
\beq
\int d\cc\int d\ww\, f_0(\cc,\ww)\ln f_0(\cc,\ww)=\int d\cc\int d\ww\, f(\cc,\ww)\ln f_0(\cc,\ww).
 \eeq
{}From Eqs.\ \eqref{2.9}--\eqref{2.6r} we obtain the corresponding entropy production as
\beq
\left.\frac{\partial s_0}{\partial t}\right|_{\text{coll}}=-\frac{3}{2}\left(\zt+\frac{\zr\Tr-2 I\bm{\Omega}\cdot\bm{\Lambda}/3}{\Tr-I\Omega^2/3}\right).
\eeq
In the case of perfectly smooth spheres ($\beta=-1$), $\bm{\Lambda}=\zr=0$ and $\zt\geq 0$, so that $\left.{\partial s_0}/{\partial t}\right|_{\text{coll}}\leq 0$ due to cooling effects \cite{K10b}. On the other hand, in the more general case of rough spheres ($\beta<1$) the Maxwellian entropy production  $\left.{\partial s_0}/{\partial t}\right|_{\text{coll}}$ does not have, in general, a definite sign.

Obviously, $s_0\geq s$ and thus the amount of missing information associated with the true distribution $f$ is upper bounded by Eq.\ \eqref{s0}. Therefore, the relevant information-theory quantity is the (local and instantaneous) \emph{relative} (or excess) entropy per particle
\beq
\Delta s(\rr,t)={s(\rr,t)-s_0(\rr,t)}=-\langle\ln R(\rr,\mathbf{v},\ww,t)\rangle, \quad R(\rr,\mathbf{v},\ww,t)\equiv \frac{f(\rr,\mathbf{v},\ww,t)}{f_0(\rr,\mathbf{v},\ww,t)}.
\label{19}
\eeq
In principle, the rigorous computation of $\Delta s(\rr,t)$ is not possible unless the solution to the Boltzmann equation \eqref{2.1} is known.
In order to circumvent this difficulty, one needs to resort to approximations.
The aim of this paper is to specialize to a homogeneous freely cooling granular gas and evaluate $\Delta s(t)$ within an \emph{approximate} scheme by assuming a truncated polynomial expansion for $R(\cc,\ww,t)$  and keeping terms at the lowest order. As will be seen below, we will mainly focus on the impact of roughness ($\beta\neq -1$) on the evolution properties of $\Delta s(t)$.

\section{Evaluation of the relative entropy}

In general, the ratio $R$ can be expanded in a complete set of orthonormal polynomials $\{\Psi_{\kk}\}$ as
\beq
R(\rr,\mathbf{v},\ww,t)=1+\sum_{\kk} A_{\kk}(\rr,t)\Psi_{\kk}(\rr,\mathbf{v},\ww,t),
\label{RR}
\eeq
where  $\kk$ denotes the appropriate set of indices and the orthonormality relation is assumed to be
\beq
\langle \Psi_{\kk}| \Psi_{\kk'}\rangle\equiv \frac{1}{n(\rr,t)}
\int d\cc\int d\ww\,  f_0(\rr,\cc,\ww,t) \Psi_{\kk}^*(\rr,\cc,\ww,t) \Psi_{\kk'}(\rr,\cc,\ww,t)=\delta_{\kk,\kk'}.
\eeq
This implies that $A_{\kk}=\langle\Psi_{\kk}^*\rangle$. Thus far, all the equations are formally exact.

Now, we first assume that $f$ is sufficiently close to $f_0$ as to apply the  \emph{linear} approximation $\langle\ln R\rangle\approx \langle R-1\rangle$. In that case, Eq.\ (\ref{19}) becomes
\beq
\Delta s(\rr,t)\approx-\sum_{\kk} |A_{\kk}(\rr,t)|^2.
\label{Deltas}
\eeq
This approximation is consistent with the negative-definite character of the relative entropy $\Delta s$.
Next, we particularize to a homogeneous and isotropic system. In that case,   $\mathbf{u}=\bm{\Omega}=0$ and the complete set of orthonormal polynomials is \cite{SVK12}
\beq
\Psi_{k_1k_2\ell}(\mathbf{v},\ww,t)=\frac{1}{\sqrt{N_{k_1k_2\ell}}}L_{k_1}^{(2\ell+\frac{1}{2})}(c^2)L_{k_2}^{(2\ell+\frac{1}{2})}(w^2)\left(c^2w^2\right)^\ell P_{2\ell}\left(\frac{\mathbf{c}\cdot\mathbf{w}}{cw}\right),\quad N_{k_1k_2\ell}\equiv \frac{\Gamma(k_1+2\ell+\frac{3}{2})\Gamma(k_2+2\ell+\frac{3}{2})}{[\Gamma(\frac{3}{2})]^2(4\ell+1)k_1!k_2!}.
\label{Psi}
\eeq
Here,
\beq
\mathbf{c}\equiv\frac{\mathbf{v}}{\sqrt{2\Tt/m}},\quad \mathbf{w}\equiv\frac{\bm{\omega}}{\sqrt{2\Tr/I}}
\eeq
are the \emph{reduced} translational and angular velocities, while
$L_n^{(\alpha)}(x)$ and $P_{2\ell}(x)$ are Laguerre (or Sonine) and Legendre polynomials, respectively.
Note that $\Psi_{k_1k_2\ell}$ is a polynomial in velocity of total degree  $2(k_1+k_2+2\ell)$. Truncation of the expansion \eqref{RR} after terms of fourth degree yields
\beqa
R(\cc,\ww,t)&\approx& 1+A_{200}(t)\Psi_{200}(\cc,\ww,t)+A_{020}(t)\Psi_{020}(\cc,\ww,t)+A_{110}(t)\Psi_{110}(\cc,\ww,t)+A_{001}(t)\Psi_{001}(\cc,\ww,t)\nn
&=&1+\frac{a_{20}}{2}\left(c^4-5c^2+\frac{15}{4}\right)+\frac{a_{02}}{2}\left(w^4-5w^2+\frac{15}{4}\right)+
a_{11}\left(c^2-\frac{3}{2}\right)\left(w^2-\frac{3}{2}\right)\nn
&&+b \left[(\mathbf{c}\cdot\mathbf{w})^2-\frac{c^2w^2}{3} \right],
\label{Sonine}
\eeqa
where the \emph{cumulants} $a_{20}$, $a_{02}$,  $a_{11}$, and $b$ are defined as
\beq
a_{20}=\frac{4}{\sqrt{30}}A_{200}=\frac{4}{15}\langle c^4\rangle-1,
\eeq
\beq
 a_{02}=\frac{4}{\sqrt{30}}A_{020}=\frac{4}{15}\langle w^4\rangle-1,
\eeq
\beq
a_{11}=\frac{2}{3}A_{110}=\frac{4}{9}\langle c^2w^2\rangle-1,
\eeq
\beq
b=\frac{2}{\sqrt{5}}A_{001}=\frac{4}{5} \left[\langle(\mathbf{c}\cdot\mathbf{w})^2\rangle-\frac{1}{3}\langle c^2w^2\rangle\right].
\eeq
The truncated Sonine expansion \eqref{Sonine} implies that Eq.\ \eqref{Deltas} becomes
\beq
\Delta s(t)\approx-\frac{15}{8}a_{20}^2(t)-\frac{15}{8}a_{02}^2(t)-\frac{9}{4}a_{11}^2(t)-\frac{5}{4}b^2(t).
\label{Deltas_a}
\eeq

{In summary, Eq.\ \eqref{Deltas_a} is obtained from the definitions \eqref{7} and \eqref{19} by assuming that the truncation  in Laguerre and Legendre polynomials, Eq.\ (\ref{Sonine}), is valid and the cumulants are taken to be small. This is the usual procedure in kinetic theory to solve the Boltzmann equation by using the Chapman--Enskog and the Grad methods. Hence the approximation that leads to the relative entropy (\ref{Deltas}) and its expression in terms of the cumulants, Eq.\ (\ref{Deltas_a}), is well justified. Notwithstanding this, there might be situations where not all the cumulants are sufficiently small \cite{SVK12}, in which case Eq.\ \eqref{Deltas_a} would be useful at a semi-quantitative level only.}

In order to obtain the time evolution of $\Delta s(t)$ we need to deal with the evolution equations for the cumulants. Taking velocity moments in both sides of the Eq.\ \eqref{2.1} (with $\nabla f=0$) it is possible to get \cite{SKS11,SVK12}
\beq
 \partial_\tau \ln \frac{\Tr}{\Tt}=\frac{2}{3}(\mu_{20}-\mu_{02}),
\label{evo1}
\eeq
\beq
  \partial_\tau\ln(1+a_{20})=\frac{4}{15}\left(5\mu_{20}-\frac{\mu_{40}}{1+a_{20}}\right),
\label{evo2}
\eeq
\beq
\partial_\tau\ln(1+a_{02})=\frac{4}{15}\left(5\mu_{02}-\frac{\mu_{04}}{1+a_{02}}\right),
\label{evo3}
\eeq
\beq
\partial_\tau\ln(1+a_{11})=\frac{4}{9}\left[\frac{3}{2}\left(\mu_{20}+\mu_{02}\right)-\frac{\mu_{22}}{1+a_{11}}\right],
\label{evo4}
\eeq
\beq
\partial_\tau\ln\left(1+a_{11}+\frac{5}{3}b\right)=\frac{4}{3}\left[\frac{1}{2}\left(\mu_{20}+\mu_{02}\right)-\frac{\mu_{b}}{1+a_{11}+\frac{5}{3}b}\right].
\label{evo5}
\eeq
Here, $d\tau=\nu_0(t)dt$, where $\nu_0(t)\equiv n\sigma^2\sqrt{{2\Tt(t)}/{m}}$ is a nominal collision frequency, and
\beq
\mu_{k_1k_2}\equiv -\frac{1}{n\nu_0}\int d\cc\int d\ww\, c^{k_1} w^{k_2} J[\cc,\ww|f,f],\quad
\mu_{b}\equiv -\frac{1}{n\nu_0}\int d\cc\int d\ww\, (\mathbf{c}\cdot\mathbf{w})^2 J[\cc,\ww|f,f],
\label{mu}
\eeq
are (reduced) collisional moments. Note that $\zt=\frac{2}{3}\nu_0\mu_{20}$ and $\zr=\frac{2}{3}\nu_0\mu_{02}$.

While Eqs.\ \eqref{evo1}--\eqref{evo5} are formally exact, they do not constitute a closed set of equations since the collisional moments are functionals of the whole distribution function $f$. On the other hand, in the spirit of the Sonine approximation \eqref{Sonine}, it is possible to insert Eq.\ \eqref{Sonine} into Eq.\ \eqref{mu} and neglect terms nonlinear in the cumulants to get
\beq
\mu_i=C_0^{(i)}(\theta)+C_{20}^{(i)}(\theta)a_{20}+C_{02}^{(i)}(\theta)a_{02}+C_{11}^{(i)}(\theta)a_{11}+C_{b}^{(i)}(\theta)b,
\quad i=20,02,11,b,
\label{Cs}
\eeq
where the twenty coefficients $C_j^{(i)}(\theta)$ depend on the temperature ratio $\theta\equiv \Tr/\Tt$, the coefficients of restitution $\alpha$ and $\beta$, and the reduced moment of inertia $\q$.
The six coefficients $C_0^{(i)}$ and $C_b^{(i)}$ for $i=20,02,b$ have been derived in Refs.\ \cite{BPKZ07,KBPZ09}, while the ten coefficients $C_{0}^{(11)}$ and $C_{20}^{(i)}$, $C_{02}^{(i)}$, and $C_{11}^{(i)}$ for $i=20,02,11$ have been derived in Ref.\ \cite{SKS11}. Finally, the four  coefficients $C_b^{(11)}$ and  $C_j^{(b)}$ for $j=20,02,11$ have been derived in Ref.\ \cite{SVK12}.
Insertion of Eq.\ \eqref{Cs} into Eqs.\ \eqref{evo1}--\eqref{evo5}, plus
further linearization with respect to the cummulants of $\mu_{40}/(1+a_{20})$, $\mu_{04}/(1+a_{02})$, $\mu_{22}/(1+a_{11})$, and $\mu_{b}/(1+a_{11}+\frac{5}{3}b)$, allows us to solve the closed set of equations for $\theta(t)$, $a_{20}(t)$, $a_{02}(t)$, $a_{11}(t)$, and $b(t)$ starting from a given initial condition. Then, the temporal dependence of $\Delta s(t)$ is obtained from Eq.\ \eqref{Deltas_a}.

\section{Results}
We have numerically solved the set of equations \eqref{evo1}--\eqref{evo5} with the Sonine approximation \eqref{Cs} for several values of $\alpha$ and $\beta$.
In all the cases the spheres are assumed to have a uniform mass distribution (i.e., $\q=\frac{2}{5}$) and the initial condition corresponds to $\theta(0)=\Tr(0)/\Tt(0)=\frac{1}{2}$, $a_{20}(0)=a_{02}(0)=a_{11}(0)=b(0)=0$, so that $\Delta s(0)=0$.

\begin{figure}
  \includegraphics[height=.32\textheight]{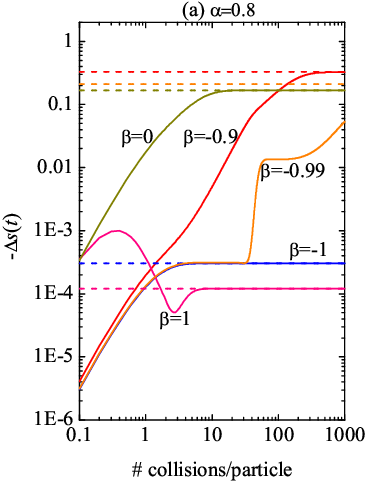}
   \includegraphics[height=.32\textheight]{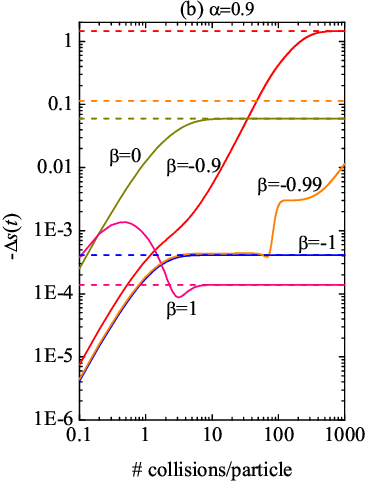}
   \includegraphics[height=.32\textheight]{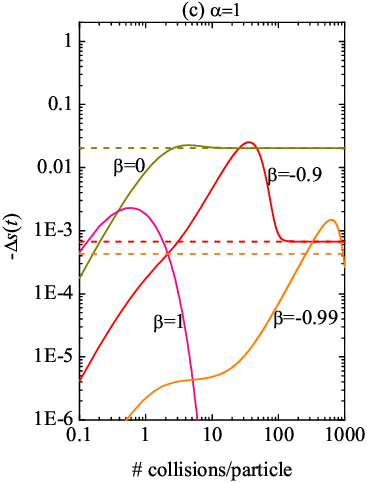}
\caption{Log-log plot of $-\Delta s(t)$ versus the number of collisions per particle for (a) $\alpha=0.8$, (b) $\alpha=0.9$, and (c) $\alpha=1$. The curves in each panel correspond to $\beta=1$, $0$, $-0.9$, and $-0.99$. Moreover,  the special case of perfectly smooth spheres ($\beta=-1$) is included in panels (a) and (b). The dashed horizontal lines represent the stationary values.
\label{fig1}}
\end{figure}

\begin{figure}
  \includegraphics[height=.25\textheight]{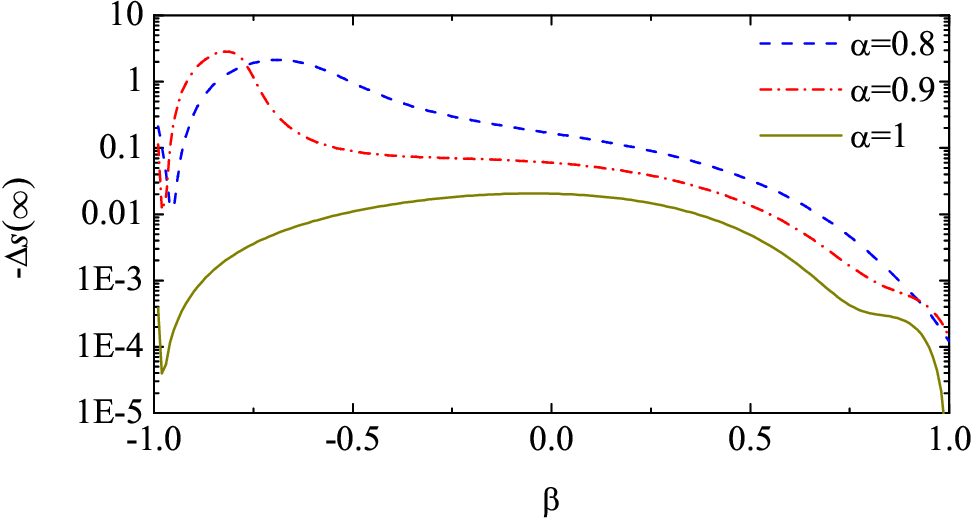}
   \caption{Plot of the stationary values $-\Delta s(\infty)$ versus $\beta$ for $\alpha=0.8$, $0.9$, and  $1$.
\label{fig2}}
\end{figure}

Figure \ref{fig1} shows $-\Delta s(t)$ for a representative number of pairs $(\alpha,\beta)$. As is commonly done, the temporal evolution is monitored not by time $t$ but by the accumulated number of collisions per particle, $\text{\#\ collisions}/\text{particle}=\sqrt{2\pi}\tau$.
A log-log plot is used in Fig.\ \ref{fig1} because of the great disparity in the orders of magnitude of both $|\Delta (s)|$ and the relaxation time for the different pairs ($\alpha,\beta)$. For perfectly rough spheres ($\beta=1$), $\Delta s$ exhibits a non-monotonic behavior and reaches its stationary value $\Delta s(\infty)$ after less than 10 collisions/particle. Note that in the conservative case of perfectly elastic and rough spheres, $(\alpha,\beta)=(1,1)$, $\Delta s(\infty)=0$. For $\beta=0$ the relaxation is slightly slower and the stationary value $|\Delta s(\infty)|$ is clearly higher than for $\beta=1$. As the spheres become less rough, a larger number of collisions is needed to activate the rotational degrees of freedom and, therefore, the relaxation period toward the stationary value becomes longer. In fact, as shown in Fig.\ 1, the stationary value corresponding to $\beta=-0.99$ has not been reached after $10^3$ collisions/particle. We have observed that, in general, the relative entropy is dominated by the contribution $-\frac{15}{8}a_{02}^2(t)$ of Eq.\ \eqref{Deltas_a}. This implies that the deviation of $s(t)$ from $s_0(t)$ is mainly due to the kurtosis of the angular velocity distribution function, the other cumulants playing a smaller role.

Figures \ref{fig1}(a) and \ref{fig1}(b) also include the evolution of $-\Delta s(t)$ for perfectly smooth spheres ($\beta=-1$). In that case, the angular velocity of each individual particle is not affected by the dynamics and thus it does not play a role different from that of a label or tag. Therefore, if $\beta=-1$ the rotational degrees of freedom are irrelevant and Eq.\ \eqref{Deltas_a} must be replaced by $\Delta s_{\text{sm}}(t)\approx-\frac{15}{8}a_{20}^2(t)$.
As can be seen from Figs.\ \ref{fig1}(a) and \ref{fig1}(b), the evolution of $\Delta s(t)$ for $\beta=-0.99$ is practically indistinguishable from that of $\Delta s_{\text{sm}}(t)$ for a transient period longer than the relaxation time of the latter quantity. However, as time progresses, the temperature ratio $\Tr/\Tt$ becomes so large that the rotational and translational degrees of freedom become strongly coupled and $\Delta s(t)$ for $\beta=-0.99$ eventually departs from $\Delta s_{\text{sm}}(\infty)$. A similar effect has been previously reported for binary mixtures \cite{S11b}.
It is important to remark that even the translational cumulant $a_{20}(\infty)$  differs for smooth spheres ($\beta=-1$) from the one obtained for quasi-smooth spheres in the limit $\beta\to -1$ \cite{SKS11}. Note that $\Delta s_{\text{sm}}(t)=0$ if $(\alpha,\beta)=(1,1)$ and so it is absent in Fig.\ \ref{fig1}(c). 

{It is important to bear in mind that $\Delta s(t)$ represents the evolution of the \emph{excess} entropy with respect to the (instantaneous) reference entropy $s_0(t)$, the latter being given by Eq.\ \eqref{s0}. While $\Delta s(t)$ is always negative, its magnitude exhibits in general a non-monotonic behavior. This means that the ``departure'' of the velocity distribution function from the two-temperature Maxwellian \eqref{2.18}, as measured by the four cumulants, exhibits a complex behavior and can increase or decrease in the course of time until  stationary values are reached. In fact, the four cumulants   have also a non-monotonic behavior, especially in the limiting cases $\alpha\rightarrow1$ and $\beta\rightarrow\pm1$ \cite{SKS11,SVK12}.}

The stationary values $-\Delta s(\infty)$ are plotted versus $\beta$ for $\alpha=0.8$, $0.9$, and  $1$ in Fig.\ \ref{fig2}. As can be seen,  $|\Delta s(\infty)|$ does not present a monotonic behavior with respect to $\beta$ and tends to increase with decreasing $\alpha$. It is also apparent that a rapid change of behavior takes place in the quasi-smooth region $\beta\approx -1$.

\section{Conclusion}

In  this paper we have studied the time evolution and stationary values of the Boltzmann entropy of a homogeneous freely cooling granular gas, relative to the maximum entropy consistent with the actual translational and rotational temperatures.
Despite the lack of an H theorem for granular gases, the quantity $\Delta s(t)$ measures the amount of missing information on the microscopic state of the gas and, consequently, represents an insightful tool to characterize the departure of the velocity distribution from a two-temperature Maxwellian.
The evaluation of $\Delta s(t)$ has been carried out by expanding the velocity distribution function in orthogonal polynomials, truncating after fourth degree, and keeping the lowest order in the coefficients. As shown by Figs.\ \ref{fig1} and \ref{fig2}, the results show a rich variety of dependencies of $\Delta s(t)$ on time,  $\alpha$, and $\beta$, including a singular behavior in the quasi-smooth limit $\beta\to -1$. We plan to undertake in the near future a comparison of the results derived in this paper with computer simulations.


\begin{theacknowledgments}
A.S.  acknowledges support from the Spanish Government through Grant No.\ FIS2010-16587 and from the Junta de Extremadura (Spain) through Grant No.\ GR10158, partially financed by FEDER  funds.
\end{theacknowledgments}



\bibliographystyle{aipproc}   

\bibliography{D:/Dropbox/Public/bib_files/Granular}

\end{document}